\documentclass[journal]{IEEEtran}
\usepackage{graphicx}
\usepackage{amsmath} 
\usepackage{amsthm}
\usepackage{amssymb}
\usepackage{eurosym} 
\usepackage{booktabs}
\usepackage[switch]{lineno}
\usepackage{xcolor}
\usepackage{color,soul}
\usepackage{acronym}
\usepackage[noadjust]{cite} 

\newcommand{\bdm}{\begin{displaymath}}
\newcommand{\edm}{\end{displaymath}}
\newcommand{\bi}{\begin{itemize}}
\newcommand{\ei}{\end{itemize}}

\newcommand{\bc}{\begin{center}}
\newcommand{\ec}{\end{center}}
\newcommand{\be}{\begin{equation}}
\newcommand{\ee}{\end{equation}}
\newcommand{\bma}{\begin{math}}
\newcommand{\ema}{\end{math}}
\newcommand{\bea}{\begin{eqnarray}}
\newcommand{\eea}{\end{eqnarray}}
\newcommand{\ba}{\begin{align}}
\newcommand{\ea}{\end{align}}
\newcommand{\bal}{\begin{aligned}}
\newcommand{\eal}{\end{aligned}}
\newcommand{\barr}{\begin{array}}
\newcommand{\earr}{\end{array}}

\begin{document}
\title{Lagrangian Decomposition based Multi Agent Model Predictive Control for Electric Vehicles Charging integrating Real Time Pricing}
%
%
%
\author{Alessandro Di Giorgio, Andrea Di Maria, Francesco~Liberati, Vincenzo Suraci, Francesco Delli Priscoli
\thanks{{The authors are with the Department of Computer, Control and  Management Engineering, 
at "Sapienza" University of Rome, Via Ariosto 25, 00185, Rome, Italy,
e-mail: }\textit{{digiorgio}}{@diag.uniroma1.it.}}
\thanks{
F. Liberati is with the SMART Engineering Solutions \& Technologies (SMARTEST) Research Center, eCampus University, Via Isimbardi 10, 22060, Novedrate CO, Italy}
\thanks{{This work is partially supported by the SAPIENZA - ATENEO 2013 "Planning and control of flexible electricity demand and generation from renewable energy sources in Smart Grids" project, no. C26A13LYTB, amd the SAPIENZA - ATENEO 2015 "SMILE - Smart MIcrogrid of eLectric Energy" Project, no. C26H15ZWK5}}}
\maketitle
\acrodef{DER}{Distributed Energy Resources}
\acrodef{DSO}{Distribution System Operator}
\acrodef{ESS}{Energy Storage System}
\acrodef{MPC}{Model Predictive Control }
\acrodef{NLP}{Non Linear Programming}
\acrodef{PLC}{Power Line Communications}
\acrodef{QCP}{Quadratic Constrained Programming}
\acrodef{QP}{Quadratic Programming}
\acrodef{RES}{Renewable Energy Sources}
\begin{abstract}
This paper presents a real time distributed control strategy for electric vehicles charging covering both drivers and grid players' needs. Computation of the charging load curve is performed by agents working at the level of each single vehicle, with the information exchanged with grid players being restricted to the chosen load curve and energy price feedback from the market, elaborated according to the charging infrastructure congestion. The distributed control mechanism is based on model predictive control methodology and Lagrangian decomposition of the optimization control problem at its basis. The simulation results show the effectiveness of the proposed distributed approach and the mutual coherence between the computed charging load curves and the resulting energy price over the time. 
\end{abstract}
\begin{IEEEkeywords}
Electric Vehicles; Utility functions; Multi Agent Model Predictive Control; Lagrangian Decomposition; Smart~Grid.
\end{IEEEkeywords}
\IEEEpeerreviewmaketitle
\section*{Nomenclature} \label{Nomenclature}
\begin{table}[h]
\small
\begin{tabular}{p{0.5cm} p{7,5cm}}
  $p$   			& EV charging power \\
  $P$   			& Power generation\\
	$\lambda$   & Shadow price\\
  $P_s$				&	Storage power \\
  $x$         & EV's state of charge\\
  $e$					&	EV's state of charge error\\
  $T_c$       & Sampling time \\
  $T_t$       & Set of time slots in the control horizon\\
\end{tabular}
\end{table}
\section{Introduction}\label{sez:intro}
\IEEEPARstart{I}{n} recent years Electric Vehicles (EVs) are receiving an increasing attention, with their number growing over the time; this is due to an increasing concern about air pollution, energy consumption, climate change and instability.
A first assessment of EVs impact has been investigated in 2009 (see \cite{EVImpact}), where it is remarked that the presence of EVs in the network will lead to a network characterized by a predominance of active elements, in addition to the traditional passive ones. From the technical point of view, another delicate effect of EVs integration is the harmonic distortions caused by these new active elements with such a high power demand. Beyond technical aspects, also 
the human-behaviour plays an important role: people tends to charge EVs during working time at their job location, then mainly at the same hours, between 10:00 AM and 6:00 PM \cite{EVImpact}); this generates high power peaks in the network, which is expected to have a detrimental effect on grid operation. 
As a matter of fact controlling the EVs charging power is becoming a real need, with some technical solutions becoming to appear in the industrial practice and academic literature.


In this paper, a decentralized \ac{MPC} strategy for EVs charging is presented. An optimization problem is solved at each sampling time with the aim of establishing the operating setpoints for the EVs chargers and an energy storage system (ESS) contributing to the coverage of the charging load; additionally the net power provided by the grid is calculated. The problem is solved in a decentralized way, resulting in a dedicated real time and dynamic local electricity market where the power schedules traded by grid and EVs agents iteratively converge to an equilibrium guaranteeing the balance between demand and supply. As customary in MPC applications, the first control sample is actually applied for operating the charging and storage infrastructure.  

The remainder of the paper is organized as follows. Section \ref{sez:soa} discusses the most modern approaches to the EVs charging problem. Section \ref{sez:scenario} presents the reference scenario used for formulating the control problem. Section \ref{sez:problem} provides the mathematical formulation and sets the basis for its solution. Section \ref{sez:procedure} presents the mathematical solution to the problem. Section \ref{sez:results} shows the results obtained in simulating the addressed scenario and finally the conclusions are drawn in section
\ref{sez:conclusions}.
\section{State of the art } \label{sez:soa}
In this section, the current state of the art for the addressed problem is presented. First of all it must be said that the problem of EV charging is truly recent and it has been of interest since few years. Even in this short time, a significant number of different approaches have been proposed.

In \cite{SoA1} there is a major attention towards the battery pack of plug-in EVs. Linear and quadratic models are used; a case study of constraints violation is performed, through battery charging schedule calculation and execution through an ordinary differential equation solver. It is demonstrated that a linear approximation is better than the quadratic one when processing boundary conditions violations; it requires less computational effort and the violation rate is almost the same of the quadratic case.

In \cite{SoA2} a "myopic" algorithm is developed to find a minimum variance strategy for EV battery charging. It does not know any information about the future and it just calculates the actual best move. This is achieved through a central unit, the Aggregator, that receives informations from EVs: it asks their required power absorption level and provides a reference value (taken within a feasible range) to follow, based on network congestion level. The convergence to optimality is reached when minimum and maximum reference values are almost the same.

In \cite{SoA3} a method based on network losses minimization is taken into account. Starting from an uncontrolled case, where power flow equations are studied in presence of voltage limits, a solver is used to find the best situation (where losses in the grid are minimized) for a given penetration level of EVs.

In \cite{SoA4} an admission control scheduling algorithm is developed. The main idea is that, when an EV arrives at a charging station, its charging session can be accepted, rescheduled or refused. The decision is made according to a greedy profit policy, which makes a comparison between the profit gained in accepting the charging session and network congestion.

In \cite{SoA5} the problem of EV charging scheduling inside a large load area is faced. Starting from a global scheduling problem (which aims to minimize costs due to EV charging), a solution is found but it is, in reality, impractical to use because of the huge number of variables and the computation time. Then, the problem is reduced to a local problem taking into account small groups of EVs. A solution is found and, by proper scaling, it is demonstrated to be close to the global one. 

In \cite{SoA6} the problem of EV charging is approached through a market-based theory. EVs have their own maximization functions in terms of power absorption, while the market has to minimize a congestion cost function. EV battery constraints and market actual maximum capacity are taken into account in order to solve the problem, and to find the optimal charging rate through a Lagrangian approach.

In \cite{SoA7} a Lagrangian approach is used. By defining user utility functions, which are dependent on power absorption, and joining them into one single maximization problem, the standard formulation application of duality theory is achieved. Constraints are grouped together and added to the primal problem in order to find a dual one. Lagrangian multipliers are updated following the anti gradient of the dual function and the optimal primal solution is found through primal problem maximization (when the dual is at its optimum).

In \cite{SoA8} the problem of EV charging is faced in two steps. The first one, consists of acquiring informations from all nodes along a travelling road and then making an estimate of the congestion level of the network; afterwards, the information on how many EVs can be charged is forwarded to single charging stations. In the second step EVs acquire data from the nearest charging station when approaching to it and then an inner algorithm takes a decision about stopping there or not, based on the battery level and the remaining distance to be covered by the EV.

In \cite{SoA9} a real-time decentralized charging algorithm is proposed. Based on a general Lagrangian approach, EVs are modeled through utility functions which are jointly taken into account in a global maximization problem. Duality theory  and the gradient descent algorithm are applied. When the latter one converges the primal problem converges to its optimal value and each EV is notified with its own optimal charging rate.

In \cite{ADMMMPC} a global maximization problem is defined and the alternating direction method of multipliers is employed to reach an optimal value for the state variables of the problem (in this case, EV charging powers). Then the method is joined with stochastic predictions made for renewable power production profiles and EVs arrival rate, in order to apply an MPC methodology and use predictions to adjust the solution of the algorithm.
	
With respect to the approaches and different formulations here reported for solving the EVs' charging problem, the proposed work is characterized as follows:
\begin{itemize}
\item First of all, the method tries to be as decentralized as possible. There is no need for a lot of data to be exchanged between the central unit and the agents. This is reached through Lagrangian relaxation and convergence of multipliers vector, which is seen as the price to pay when buying power.
\item Then, the vectors of variables must be of small length, because otherwise the computational time becomes higher. In order to reach this goal, only one vector is broadcast to all agents and it is, indeed, the price vector: it drives user choices in such a way that they can decide on their own how much power to ask, without any knowledge of the network (except for price);
\item MPC approach is used. It is effective in order to keep the system, at each time instant, on the right evolution towards global equilibrium and single agent optimization. Most of previous case studies, in fact, were only focused on real-time implementation and not on preview about what could happen.
\item A storage element is introduced in problem formulation, in order to help the algorithm in distributing peak requests that otherwise would cause instability and faults in charging station.
\end{itemize}
With this in mind, the proposed EV charging reference scenario and strategy is presented in the following sections.
\section{Reference scenario} \label{sez:scenario}
%
\subsection{Actors and Systems}
The scenario addressed in this paper is composed of these entities (actors and systems).
\subsubsection{Charging Stations}
The Charging Station is the main plant of the scenario. It is actually the place where all the electric vehicles come asking for powers and it has the responsibility of distributing power to them. It is equipped with some access points where the EVs can plug-in their chargers and proceed in acquiring power.
It acts like a middleware between providers and customers, since it has the duty of taking power from the formers and distribute it to the latter.
\subsubsection{Distribution System operator (DSO)} The Distributor System Operator is the entity in charge of providing power to charging station. It is equipped with a storage element that can lend free power when needed. It is directly connected to the charging station and it communicates to this one the quantity of power that it has determined to sell (at current price).
\subsubsection{Drivers} They need to have their battery reach a desired charge level, before their departure time.
\subsubsection{Aggregator} It is the entity that has the duty of determining power price. It receives both requests of EVs and DSO, then it proceeds in determining the best price; it does it by iteratively asking power curves to the other actors and adjusting price. It is usually mounted inside the charging station (as a control logic) but it can be external to it; in this second case it should be directly connected to all the other actors.
\subsection{Use Case}
Here a short description of entities interaction is given.
During a certain amount of time (it may be a day, a week, or an hour) an unknown number of electric vehicles (drivers) come to a charging station. Their arrival time is completely random and there will be overlapping requests (in a real case, a lot of EVs can come charging at the same time, as discussed in introduction). They connect asking for powers and communicate their desired power curves, which are elaborated based on current predictions of power price. Charging station asks to the Distributor System Operator how much power it can sell at the actual predicted price and it obtains the offer curve. The Aggregator receives this data and, by iteratively updating price levels and collecting curves from EVs and DSO, it reaches an optimal price value to be used for all the predicted time (thus, the optimal power curves).
After this, the DSO provides required power and the Charging Station distributes it to EVs.
This whole process is repeated for every time slot of the day.
\section{Control problem formalization} \label{sez:problem}
Some key points are set here, in order to better specify the situation:
\begin{itemize}
\item Time is split into time slots. The actual time is named "t". The method is developed to work in any time-scale.
\item There is a prediction time window $T_t$. It is the set of time slots during which actual EVs will be active. It is composed of N slots, and ranges from actual time t up to $N-1$ slots in the future. $T_t = t, t+1, \cdots, t+N-1$.\\
In order to calculate $T_t$, two methods can be used:
\begin{itemize}
\item[1] Time window has fixed length N, and all the EVs must ask to charge their battery maximum before N time slots.
\item[2] Time window goes from actual time "t" up to the maximum time (that can be waited for charging) between \textbf{actual} EVs.
\end{itemize}
\item Since there are predictions (due to MPC), when referring to predicted quantities (and curves) each variable is of the kind $x(\tau/t)$, which means that the variable value is previewed at time $\tau$ with respect to actual time t, where $\tau$ belongs to Time Window $T_t$.
\item The number of EVs changes over time, because they can arrive or end their charging process. So the number of EVs active at time t is $R_t$. 
\item Prediction is given only for power curves. The number of EVs at time t ($R_t$) is the same along all the actual predicted time window $T_t$. At next time slot, if an EV arrives, time window is changed and so $R_t$ value.
\item There is only one power distributor, which is assumed s well to operate a storage element. Power withdrawn from storage does not have a cost.
\end{itemize}
To every EV a single Utility function is assigned, which is a function of the power exchanged with the charging station, namely $U(p_r(\tau/t))$. This utility function has three important properties:
\begin{itemize}
\item It is monotonically non-decreasing: this is due to the fact that the level of satisfaction of each user grows up with the level of power consumption.
\item It is concave: this is due to the fact that satisfaction of the user can have saturation.
\item It is continuous.
\end{itemize}
Each EV tries to maximize its own utility function and demands the charging station as much power as possible. Its request in terms of power absorption is limited by two important factors: the maximum physical power capability of the EV and the quantity of power to absorb in order to reach desired state of charge. The maximization is done on all the temporal window that each EV can see.
\begin{equation}
\max_{p_r(\tau/t) \in I_r(t)}\sum_{\tau \in T_t} U_r(p_{r}(\tau/t))
\end{equation}
Subject to the boundary conditions on EV power absorption limits
\begin{equation}
p_r^{min} < p_r(\tau/t) < p_r^{max} \text{ , } \forall r \in R_t \text{ , } \forall \tau \in T_t 
\end{equation}
and to state of charge error conditions
\begin{eqnarray*}
e_r(\tau+1/t)=e_r(\tau/t)-(1 - \xi)T_c p_r(\tau/t) \text{ , } & \forall \tau \in T_t \text{,} r \in R_t  \\
e_r(t_f/t)=0 \text{ , } & \forall r \in R_t \\
e_r(t/t)=e(t) \text{ , } & \forall r \in R_t
\end{eqnarray*}
where $\xi \in [0,1]$.
Power production , instead, tries to sell the quantity of power that it is more convenient. Thus its utility function tries to minimize costs of production (that, in current scenario, is equal to maximize profits).

Its utility function depends on the quantity of power that it has to produce at every time, i.e. $C_l(P_l(\tau/t))$. It has the following properties:
\begin{itemize}
\item It is monotonically non-decreasing: the cost of providing a certain amount of power should be increasing with the level of energy capacity.
\item It is convex.
\item It is continuous.
\end{itemize}
In a more explicit form\\
\begin{equation}
\min_{P_l(\tau/t) \in I_l{t}} \sum_{\tau \in T_t}C_l(P_l(\tau/t))
\end{equation}
subject to
\begin{equation}
P_l^{min} < P_l(\tau/t) < P_l^{max} \text{ , } \forall \tau \in T_t 
\end{equation}
The storage element provides "free" energy to power production when needed; it has an utility function that tries to charge back power when it can and tries to keep the state of charge as close as possible to a reference value.
It can be formulated as
\begin{equation}
\sum_{j=t}^{t+N-1} (x_s(t) - \sum_{i=t}^j P_s(i/t) \delta_s T_c - x_{ref})^2
\end{equation}
with boundary condition of
\begin{equation}
P_s^{min} < P_s(\tau/t) < P_s^{max} \text{ , } \forall \tau \in T_t
\end{equation}
By summing up all the utility functions of all the agents (for all the previewed time) a \textbf{Global Utility Function} is obtained.
\begin{align}\label{eq:globalUtility}
\max_{\begin{matrix}p_{r}(\tau/t) \in I_{r}(t) \\
 P_l(\tau/t) \in I_l(t) \\ P_s(\tau/t) \in I_s(t) \end{matrix}} & \sum_{r \in R_t} \sum_{\tau \in T_t} U_r(p_{r}(\tau/t)) \nonumber \\
 & - \sum_{\tau \in T_t}C_l(P_l(\tau/t)-P_s(\tau/t)) \nonumber \\
 & - \sum_{j=t}^{t+N-1}(x_s(t)- \sum_{i=t}^{j}P_s(i / t) \delta_s T_c -x_{ref})^2
\end{align}
With 3 types of agents acting at the same time, the only fair solution to the problem is to give every user the maximum possible degree of satisfaction with the same priority of the others. This is actually trying to maximize the \textbf{Social Welfare}.
This must be done while considering the physical constraints of the problem
\begin{equation}
\label{C1}
\sum_{r \in R_t}p_r(\tau/t)=P_l(\tau/t) \text{ , } \forall \tau \in T_t
\end{equation}

\section{Decentralized solving procedure} \label{sez:procedure}
In order to accomplish the task of solving the maximization problem, one possible formulation is to use the global constraints with \textbf{Lagrangian Multipliers} and to introduce them inside the maximization problem, as in standard Lagrangian Theory.
Local constraints are of no concerns towards the convergence of the algorithm, because they only limit some quantities, but they do not influence global multipliers.

A Lagrangian variable is needed, in order to have a unique key quantity that is shared among all the agents and can help regulating price of energy.
A new vector is required and it is
$\vec{\lambda(t)} = [\lambda(t/t), \lambda(t+1/t), \cdots , \lambda(t+N-1/t)]$.
By pre-multiplying \ref{C1} with $\vec{\lambda(t)}$, the standard formulation of \cite{dualityTheory} is achieved.
\begin{align}
L(\vec{p_r(t)}, \vec{P_l(t)},\vec{P_s(t)}, \vec{\lambda(t)}) = \sum_{r \in R_t} \sum_{\tau \in T_t} U_r(p_{r}(\tau/t)) +  \nonumber \\
 - \sum_{\tau \in T_t}C_l(P_l(\tau/t)-P_s(\tau/t)) + \nonumber \\
 - \sum_{j=t}^{t+N-1}(x_s(t)- \sum_{i=t}^{j}P_s(i / t) \delta_s T_c -x_{ref})^2 + \nonumber \\
 -\sum_{\tau \in T_t}(\lambda(\tau/t) (\sum_{r \in R_t}p_r(\tau/t) - P_l(\tau/t)))
\end{align}
The Lagrangian can be split into $\#R_t+1$ separate problems: $\#R_t$ problems, for every EV, and one for distributor.
\begin{align}
L(\vec{p_r(t)}, \vec{P_l(t)},\vec{P_s(t)}, \vec{\lambda(t)}) = \nonumber \\
 \sum_{r \in R_t} \sum_{\tau \in T_t}( U_r(p_{r}(\tau/t)) -\lambda(\tau/t)p_r(\tau/t))+  \nonumber \\
 - \sum_{\tau \in T_t}(C_l(P_l(\tau/t)-P_s(\tau/t)) - \lambda(\tau/t)P_l(\tau/t)) )+ \nonumber \\
 - \sum_{j=t}^{t+N-1}(x_s(t)- \sum_{i=t}^{j}P_s(i / t) \delta_s T_c -x_{ref})^2
\end{align}
They can be solved separately. So the \textbf{single EV problem} is:
\begin{align}\label{eq:singleEV}
\left\{\begin{matrix}
\max_{\vec{p_r(t)}} \sum_{r \in R_t} \sum_{\tau \in T_t}( U_r(p_{r}(\tau/t)) -\lambda(\tau/t)p_r(\tau/t)) \\
\text{subject to constraints a,b,c,d}
\end{matrix}\right.
\end{align}
where (a),(b),(c),(d) are
\begin{itemize}
\item[a] $p_r^{min} < p_r(\tau/t) < p_r^{max} \text{ , } \forall r \in R_t \text{ , } \forall \tau \in T_t$ 
\item[b] $e_r(\tau+1/t)=e_r(\tau/t)-(1 - \xi)T_c p_r(\tau/t) \text{ , }  \forall \tau \in T_t \text{,} r \in R_t$
\item[c] $e_r(t_f/t)=0 \text{ , }  \forall r \in R_t$
\item[d] $e_r(t/t)=e(t) \text{ , }  \forall r \in R_t$
\end{itemize}
The \textbf{DSO problem} is:
\begin{align}\label{eq:singleDSO}
\left\{\begin{matrix}
\min_{\vec{p_r(t)}} \sum_{\tau \in T_t}(\lambda(\tau/t)P_l(\tau/t))  - C_l(P_l(\tau/t)-P_s(\tau/t)) \\
- (x_s(t)- \sum_{i=t}^{j}P_s(i / t) \delta_s T_c -x_{ref})^2 \\
\text{s.t. }\\
P_l^{min} \leq P_l(\tau/t) \leq P_l^{max} \forall \tau \in T_t \\
P_s^{min} \leq P_s(\tau/t) \leq P_s^{max} \forall \tau \in T_t
\end{matrix}\right.
\end{align}
By solving \ref{eq:singleEV} and \ref{eq:singleDSO}, the Dual function D is found.
\begin{align}
\min_{\vec{\lambda(t)}} D(\vec{\lambda(t)}) 
\end{align}
where
\begin{align}
D(\vec{\lambda(t)}) = \max_{\begin{matrix}
\vec{p_r} \in I_r(t) \\
\vec{P_l} \in I_l(t) \\
\vec{P_s} \in I_s(t)
\end{matrix}} 
L(\vec{p_r(t)}, \vec{P_l(t)},\vec{P_s(t)}, \vec{\lambda(t)}) 
\end{align}
Then, in order to find the optimal value for the dual problem, a variation of Gradient Projection algorithm is used. The Lagrangian multipliers are updated with the projected anti-gradient of D.
\begin{align}\label{eq:lambdaEvo}
\vec{\lambda_{k+1}(t)} = \max(\vec{\lambda_k(t)}-\gamma(\nabla D),0)
\end{align}
where\\
\begin{align}
\nabla D =  \vec{P}_l(t) -\sum_{r \in R_t} \vec{p}_r(t)
\end{align}
Thus the final algorithm is
\textbf{Solution Algorithm}
\begin{itemize}
 \item[1] At time slot t of the day, calculate the number of active EVs ( $R_t$ ).
\item[2] For each of the active EV ask its departure time, and set the maximum value among them as Time Horizon. This will be the limit of the optimization window ($T_t$).
\item[3] During same time slot t, execute this cycle until exit condition is met:
\begin{itemize}
 \item[a] Iteration k=0, take price of previous time slot as starting price and set all
prices along time window $T_t$ the same.
\item[b] Each vehicle solves its own optimization problem (\ref{eq:singleEV}), from actual time up to its departure time, calculating $\vec{p}_r(t)$.
\item[c] The information of every $\vec{p}_r(t)$ is forwarded to the charging station and the aggregator sums them together. After that the DSO solves its own optimization problem (\ref{eq:singleDSO}) to calculate the power which can give away ($\vec{P}_l(t)-\vec{P}_s(t))$.
\item[d] Price vector is updated following the antigradient of the dual problem (Eq \ref{eq:lambdaEvo}).
\item[e] If the exit condition is met, price vector is the optimum, so exit this inner loop and go to 4. Otherwise, set $k\leftarrow k+1$ and repeat from b to e, using new prices and new powers as starting values for next iteration.
\end{itemize}
\item[4] Once 3 has ended, the price value obtained is the optimal price at current time slot t, and the first value of the vector is taken as price update for next time slot "t+1". Powers are extracted by solving EV and DSO problem and they are the optimum curves to follow.
\item[5] Set $t \leftarrow t+1$ (i.e. go ahead of one time slot 1) and repeat steps from 1 to end.
\end{itemize}
\section{Simulation Results} 
\label{sez:results}
\begin{figure}
\centering
\includegraphics[width=\linewidth]{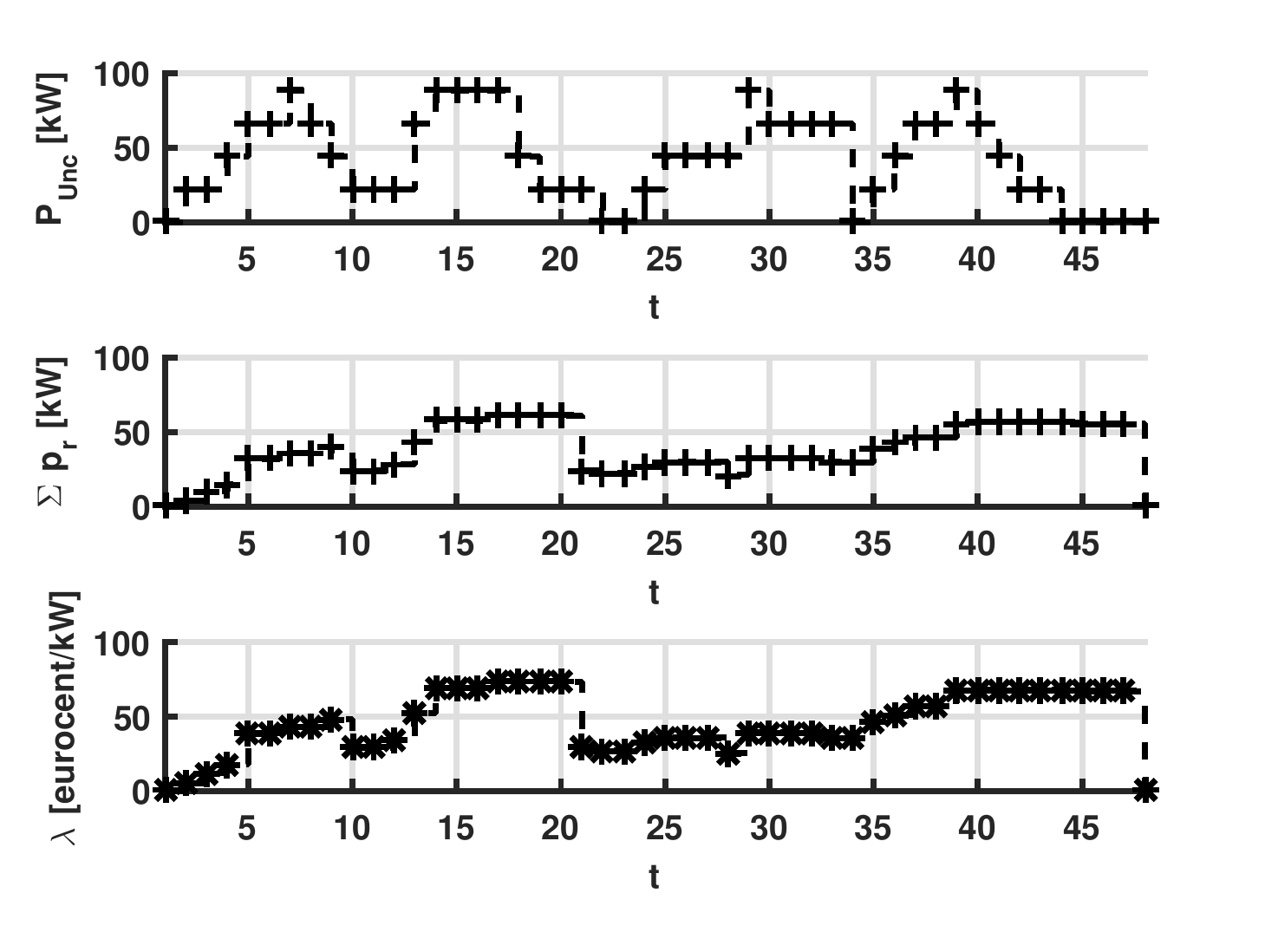}
\caption{Case Without Storage}
\label{fig:DailyNoStorage}
\end{figure}

Simulations have been performed using an iMac 21.5, Intel Core i5, 2,7 GHz,  12 GB RAM 1333 Mhz DDR2 computer, running Apple OSX  10 (v.11). The control framework has been built in Matlab 64 bit, and the \ac{MPC} problem has been solved at each iteration by using Matlab built-in solver.
The base case is a half day scenario (12 hours), with time split into 15 min time slots, for a total of 48 time slots. EVs' arrivals and departures have been randomly generated.
Simulation parameters are specified in table (\ref{table:SimScenario4.3}).
\begin{table}[h]
\centering
\caption{Load Area Simulation Scenario 3}
\label{table:SimScenario4.3}
\begin{tabular}{|lll|}
\hline
\# EV & Variable & 
$\begin{matrix}$Logarithmic Utility Function
$ \\ U_r(p_{r}(t))\quad \forall r \in R \\
I_r(t)$ for each consumer $ \\
p_{r}(t) \in [0;22] kW \\
$Weights $\\
w_{r}(t)=10; \quad for \text{EVs} \quad 1 \text{ to } 20
\end{matrix}$ 
\\ \hline
\# Energy Sources & $\quad$ 1 & 
$\begin{matrix}$DSO cost function$ \\ 0.06(P_l(t)-P_s(t))^2+\\ +0.9(P_l(t)-P_s(t))\end{matrix}$\\  \hline
Initial Prices &  & $16 \quad $\euro  cent /kWh
 \\ \hline
$P_l^{max}(t)$	 &  & $100 kW $ \\ \hline
\# Storage & $\quad$ 1 & 
$\begin{matrix}$Storage cost function $ \\ (x_s(t_0) - \sum_{i=t}^{\tau}P_s(i/t)\delta_s T_c-x_{ref})^2 \\
= (x(t) - x_{ref})^2\end{matrix}$\\  \hline
$P_s^{min}$ & & $-100 kW$ \\ \hline
$P_s^{max}$ & & $100 kW$ \\ \hline
$x_s(t_0)=x_{ref}$ & & $100 kWh$ \\ \hline
\end{tabular}
\end{table}
First of all, in Fig. \ref{fig:DailyNoStorage}, an earlier simulation has been performed without any storage element. 

This is a development of the basis theory formalized in \cite{med}, where a similar approach has been set only for a single time slot. The first plot is the uncontrolled power profile, which is obtained when the EVs arrive at the station and  they charge their battery at maximum power. Those peaks are the ones to be avoided.

The second one is the evolution of demand curve along the day, once the control algorithm proposed in this paper has been applied; of course, since there is a balance equation between demand and offer, the demand curve exactly matches the offer curve. The third one is the daily evolution of price. It follows the same shape of offer curve.

\begin{figure}
\centering
\includegraphics[width=\linewidth]{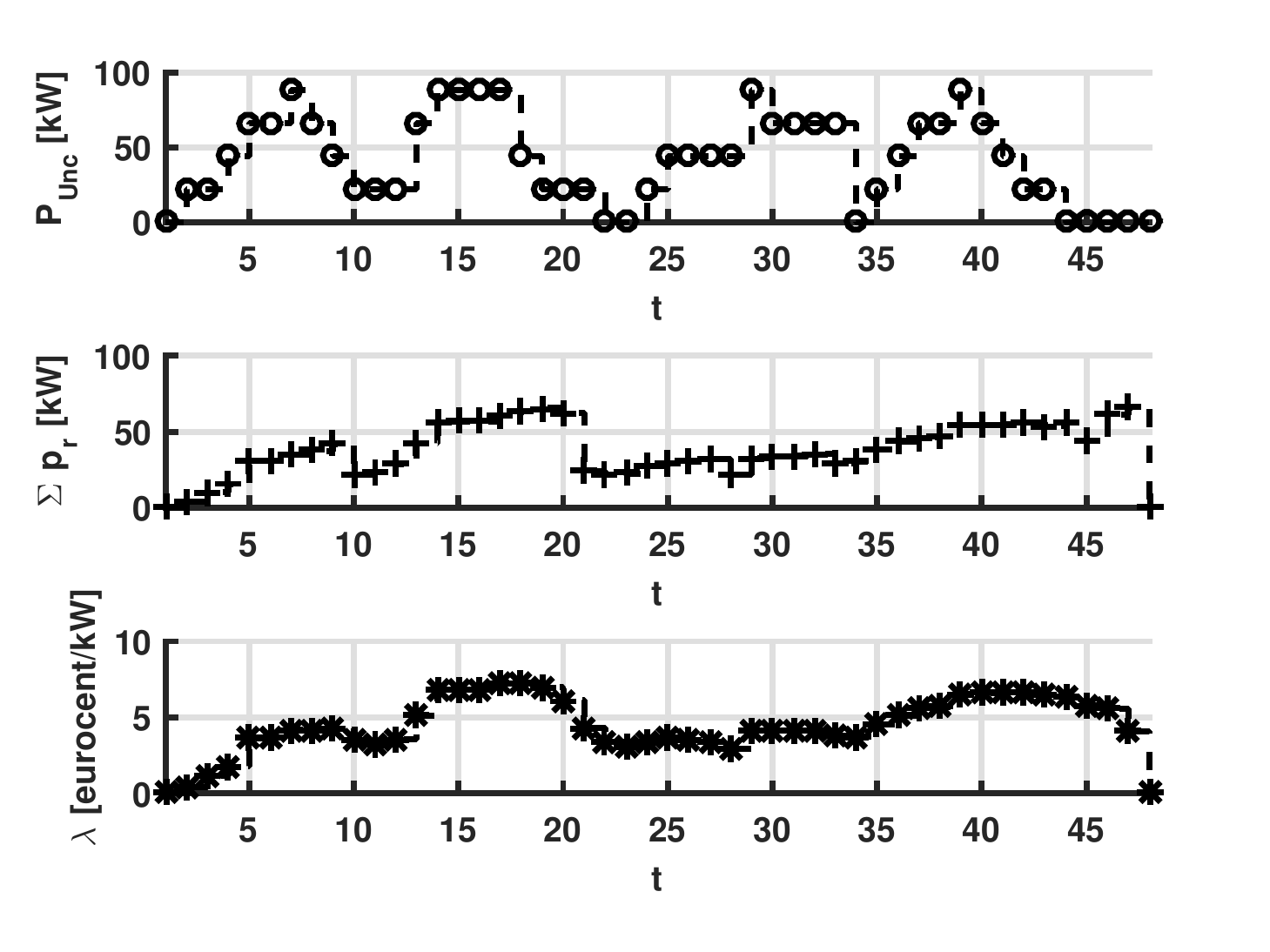}
\caption{Powers and Price}
\label{fig:DailyStorage}
\end{figure}

\begin{figure}
\centering
\includegraphics[width=\linewidth]{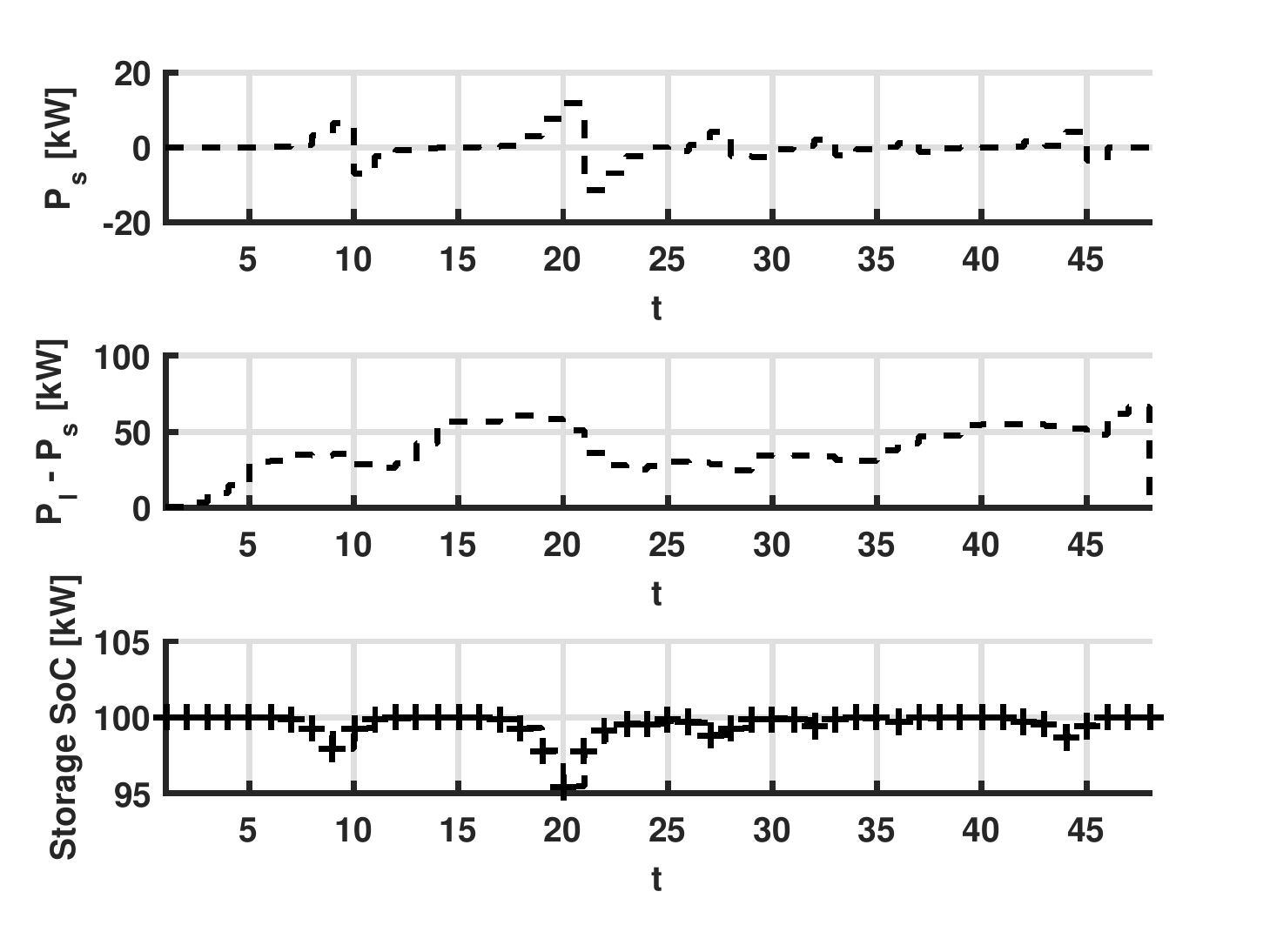}
\caption{Agregated net power flow and storage behaviour.}
\label{fig:DailySoc}
\end{figure}

When a storage element is introduced (Fig. \ref{fig:DailyStorage}) demand and offer curves stay almost the same of the case of no storage($2^{nd}$ plot), but the evolution of prices changes (see third plot). 
The evolution of prices is heavily reduced, more or less of 1/5, with respect to before and its shape is smoother, with less variations. Moreover, one can note that price has a different shape with respect to power offer curve.
In Fig. \ref{fig:DailySoc} the reason behind the new price curve is explained: since the storage element is lending "free" power to production side (i.e. there is no cost of production for $P_s$), the quantity of power that is produced is modified accordingly, in order to keep the Utility Cost Function of the producer at its optimal value. 
Since price follows production shape, it goes alongside with $P_l-P_s$, which is the net produced power. Storage is employed to retain the maximum possible reward in producing as long as it does not discharge too much the storage element.
\section{Conclusions} \label{sez:conclusions}
In this paper a decentralized MPC approach for EVs smart charging in a load area has been presented. The problem has been considered first from a global perspective, building a standard Lagrangian formulation with a set of balance equations as the unique constraints. Then a decentralization step has been performed, through decomposition of the global problem into small subproblems, one for each EV and one more for the DSO, with their own local constraints. Then the dual problem has been introduced. Finally, a decentralized MPC algorithm has been defined in order to control power curves evolution: the method has been demonstrated to achieve convergence for variables of both primal and dual problems.  Simulations have been performed for a case of an half day scenario in a small area, equipped with a storage element. They showed the effectiveness of the method when using the storage  and the fast convergence rate towards network balance and users' satisfaction. Some considerations have been done regarding the shadow price variable and the storage, assessing the possible role of the former and the latter's contribution towards the smoothness of offered power curve.

\bibliographystyle{IEEEtran}
\bibliography{References}

\begin{thebibliography}{10}
\providecommand{\url}[1]{#1}
\csname url@samestyle\endcsname
\providecommand{\newblock}{\relax}
\providecommand{\bibinfo}[2]{#2}
\providecommand{\BIBentrySTDinterwordspacing}{\spaceskip=0pt\relax}
\providecommand{\BIBentryALTinterwordstretchfactor}{4}
\providecommand{\BIBentryALTinterwordspacing}{\spaceskip=\fontdimen2\font plus
\BIBentryALTinterwordstretchfactor\fontdimen3\font minus
  \fontdimen4\font\relax}
\providecommand{\BIBforeignlanguage}[2]{{%
\expandafter\ifx\csname l@#1\endcsname\relax
\typeout{** WARNING: IEEEtran.bst: No hyphenation pattern has been}%
\typeout{** loaded for the language `#1'. Using the pattern for}%
\typeout{** the default language instead.}%
\else
\language=\csname l@#1\endcsname
\fi
#2}}
\providecommand{\BIBdecl}{\relax}
\BIBdecl

\bibitem{EVImpact}
C.~Yang and R.~McCarthy, ``Electricity grid: Impacts of plug-in electric
  vehicle charging,'' \emph{University of California}, 2009.

\bibitem{SoA1}
O.~Sundstrom and C.~Binding, ``Optimization methods to plan the charging of
  electric vehicle fleets,'' \emph{IBM Research}, 2010.

\bibitem{SoA2}
Q.~Li and T.~Cui, ``On-line decentralized charging of plug-in electric vehicles
  in power systems,'' \emph{arXiv, Cornell University Library}, 2011.

\bibitem{SoA3}
D.~S. A.~T. LE and M.~O. VAZQUEZ, ``Scheduling charging of electric vehicles
  for optimal distribution systems planning and operation,'' \emph{21st
  International Conference on Electricity Distribution}, 2011.

\bibitem{SoA4}
Y.~J. Shiyao~Chen, , and L.~Tong, ``Deadline scheduling for large scale
  charging of electric vehicles with renewable energy,'' \emph{School of
  Electrical and Computer Engineering Cornell University}, 2012.

\bibitem{SoA5}
Y.~He and B.~Venkatesh, ``Optimal scheduling for charging and discharging of
  electric vehicles,'' \emph{IEEE Transactions on Smart Grids}, 2012.

\bibitem{SoA6}
J.~Hu and ShiYou, ``Coordinated charging of electric vehicles for congestion
  prevention in the distribution grid,'' \emph{IEEE Transactions on Smart
  Grids}, 2013.

\bibitem{SoA7}
C.~R. Omid~Ardakanian and S.~Keshav, ``Distributed control of electric vehicle
  charging,'' \emph{e-Energy '13 Proceedings of the fourth international
  conference on Future energy systems}, 2013.

\bibitem{SoA8}
Z.~Q. Azwirman~Gusrialdi and M.~A. Simaan, ``Scheduling and cooperative control
  of electric vehicles’ charging at highway service stations,'' \emph{the
  53rd IEEE Conference on Decision and Control}, 2014.

\bibitem{SoA9}
J.~Rivera and C.~Goebel, ``A distributed anytime algorithm for real-time ev
  charging congestion control,'' \emph{Conference: Proceedings of the 2015 ACM
  Sixth International Conference on Future Energy Systems, At Bangalore,
  India}, 2015.

\bibitem{ADMMMPC}
D.~O. Trudie~Wang and H.~Kamath, ``Dynamic control and optimization of
  distributed energy resources in a microgrid,'' \emph{Smart Grid, IEEE
  Transactions on (Volume:6 , Issue: 6 )}, 2015.

\bibitem{dualityTheory}
R.~Freund, ``Duality theory of constrained optimization,'' \emph{MIT courses},
  2004.

\bibitem{med}
A.~Q. Andrea~Mercurio, Alessandro Di~Giorgio, ``Distributed control approach
  for community energy management systems,'' \emph{MED 2012}, 2007.

\end{thebibliography}
\end{document}